\documentclass[submission,copyright,creativecommons]{eptcs}
\providecommand{\event}{SOS 2011}
\usepackage{amssymb, amsmath}
\usepackage[mathletters]{ucs}
\usepackage[utf8x]{inputenx}
\usepackage{listings}
\usepackage{mathpartir}
\usepackage{color}
\usepackage{breakurl}
\usepackage{stmaryrd}

\newtheorem{property}{Property}
\newtheorem{lemma}{Lemma}

\definecolor{dkblue}{rgb}{0,0.1,0.5}
\definecolor{lightblue}{rgb}{0,0.5,0.5}
\definecolor{dkgreen}{rgb}{0,0.4,0}
\definecolor{dk2green}{rgb}{0.4,0,0}
\definecolor{dkviolet}{rgb}{0.6,0,0.8}

\newcommand{\LT}[3]{#1 \xrightarrow{#2} #3}

\newcommand{\isdef}{\,\hat{=}\,}

\newcommand{\on}[1]{\textnormal{\textsf{#1}}}

\newcommand{\seq}[2]{\on{seq}\,(#1,\,#2)}
\newcommand{\onskip}{\on{skip}}
\newcommand{\letin}[2]{\on{block}\,(#1,\,#2)}

\newcommand{\Cmd}{\mathbf{Cmd}}
\newcommand{\Exp}{\mathbf{Exp}}

\newcommand{\Dcl}{\mathbf{Dcl}}

\newcommand{\Pcd}{\mathbf{Pcd}}

\newcommand{\Prm}{\mathbf{Prm}}

\newcommand{\Object}[1]{#1}

\title{Formal Component-Based Semantics}
\author{Ken Madlener
\institute{Institute for Computing and Information Sciences (iCIS),\\
Radboud University Nijmegen, The Netherlands}
\email{k.madlener@cs.ru.nl}
\and Sjaak Smetsers
\institute{Institute for Computing and Information Sciences (iCIS),\\
Radboud University Nijmegen, The Netherlands}
\email{s.smetsers@cs.ru.nl}
\and Marko van Eekelen
\institute{Institute for Computing and Information Sciences (iCIS),\\
Radboud University Nijmegen, The Netherlands}
\institute{School of Computer Science, Open University of the Netherlands}
\email{m.vaneekelen@cs.ru.nl}
}
\def\titlerunning{Formal Component-Based Semantics}
\def\authorrunning{Madlener, Smetsers \& van Eekelen}
\date{today}

\begin{document}
\maketitle

\begin{abstract}
One of the proposed solutions for improving the scalability of semantics of programming languages is Component-Based Semantics, introduced by Peter D. Mosses. It is expected that this framework can also be used effectively for modular meta theoretic reasoning. This paper presents a formalization of Component-Based Semantics in the theorem prover \textsc{Coq}. It is based on Modular SOS, a variant of SOS, and makes essential use of dependent types, while profiting from type classes. This formalization constitutes a contribution towards modular meta theoretic formalizations in theorem provers. As a small example, a modular proof of determinism of a mini-language is developed.
\end{abstract}

\section{Introduction}\label{sec:introduction}


Theorem prover formalization of programming language meta theory and semantics receives a lot of attention. Most notably, the {\sc PoplMark} Challenge~\cite{ayd05:poplmark} calls for experiments on verifications of meta theory and semantics using proof tools. One of the main issues that programming language formalizations have to cope with is the lack of reusability of existing work. Many programming languages have language constructs in common, but often have (slight) differences in their precise semantics (e.g. assignments in {\sc C} versus assignments in {\sc Java}).

Component-Based Semantics, introduced by Peter D. Mosses, aims to resolve this reusability issue by constructing language descriptions from combinations of basic abstract constructs~\cite{Mosses09}. Basic constructs are supposed to have a fixed meaning and be language-independent. As an example, the basic construct of conditional expressions should not depend on whether the expressions may have side-effects or not, terminate abruptly or even interact with other processes. One could even go as far as creating a repository of constructs that may be freely combined to build new languages. This repository is therefore necessarily open-ended, enabling users to add newly discovered basic constructs.

Modular Structural Operational Semantics (MSOS)~\cite{Mosses04}, a variant of SOS, provides an adequate framework for the independent description of language components~\cite{Mosses09}. MSOS was designed to address the lack of reusability of SOS rules: \emph{every} auxiliary entity used in a rule, such as an environment or a store, needs to be threaded through \emph{all} rules of the language. MSOS provides a way to automatically propagate unmentioned entities between the premise(s) and conclusion of a rule, enabling the reuse of rules in different languages. SOS is very suitable for the formalization of languages and has therefore been widely adopted by the theorem prover community. MSOS has so far received less attention. 

This paper proposes a formalization of Component-Based Semantics based on MSOS in the theorem prover {\sc Coq}~\cite{CoqManualV83}.\footnote{The source can be obtained at \url{http://www.cs.ru.nl/~kmadlene/fcbs.html}.} Our main contribution is a way to constructively formalize programming language semantics: basic constructs can be developed in separate {\sc Coq} files, which may be verified independently. The formalization has been tested by building a small repository of constructs. Moreover, it is possible to equip the constructs with small proofs that can be used to construct larger proofs of properties holding for a full language. For this reason, we shall use the term \emph{component} instead of \emph{construct} in this paper. Our formalization supports meta theoretic reasoning about a programming language, but does not support reasoning about the format of MSOS rules.

The formalization follows the original design of MSOS in its use of arrows of a category for the auxiliary entities (encapsulated in labels) appearing in the transition rules. A very elementary level of knowledge about category theory and a modest amount of familiarity with theorem proving is required to read this paper. Our formalization makes essential use of dependent types to formalize the labels in MSOS, and profits from {\sc Coq}'s support for type classes. Each component is represented by a parametrized so-called {\sc Coq} section. To define a full language, it is sufficient to enumerate its components. The correct instantiation of the corresponding parameters can in principle be performed automatically by {\sc Coq}'s powerful type system.

\section{Component-Based Semantics}\label{sec:cbmsos}

We illustrate the description of programming languages in terms of basic abstract constructs by means of a while-loop example taken from~\cite{Mosses09}. Depending on what concrete language is being analyzed, a standard command such as $\on{while}$ may have different interpretations. For example, if the language includes a $\on{break}$ command that abruptly terminates the program throwing a particular exception, then the description of $\on{while}$ should include the handler for that exception. We assume that $Cmd \llbracket\,\_\,\rrbracket$ and $Exp\llbracket\,\_\,\rrbracket$ are functions mapping concrete expressions to abstract expressions of $\Cmd$ and $\Exp$, respectively. Below, $\on{cond-loop}$ is a simple while-loop that takes an expression and a command, and propagates abrupt termination. The other constructs involved can be found in Table~\ref{fig:universal}. The description is then:

\begin{center}
\begin{tabular}{r@{}c@{}l}
$\begin{array}{@{}c@{}}Cmd \llbracket \on{while}\ (E)\ C \rrbracket\\{}\end{array}$ & $\begin{array}{c}=\\{}\end{array}$ &
$\begin{array}{@{}l@{}l}
  \on{catch}( & \on{cond-loop}(Exp \llbracket E \rrbracket,\ Cmd \llbracket C \rrbracket),\\
  &\on{abs}(\on{eq}(\on{breaking}),\ \on{skip}))
 \end{array}$
\vspace{0.2cm}\\
$Cmd \llbracket \on{break} \rrbracket$ &$=$& $\on{throw}(\on{breaking})$
\end{tabular}
\end{center}

\noindent
A simple extension is a while-loop that handles $\on{continue}$ commands. To describe such while-loops, all that is needed is to change the above example in such a way that $Cmd \llbracket C \rrbracket$ is encapsulated by a $\on{catch}$ construct. Table~\ref{fig:universal} contains some possible constructs, which are used as examples throughout the rest of this paper. An example of an open-ended repository containing more constructs can be found in e.g.~\cite{Mosses09}.

An important facet of Component-Based Semantics is that the construct repositories ideally contain no redundancy. If two basic constructs with different names have the exact same semantics, then one of them should be discarded. Moreover, if a construct can be expressed purely in terms of existing basic constructs, then this construct should also be discarded. A repository therefore essentially describes a universal language that can be used to define the semantics of a concrete language in question. This universal language provides a fixed name for each basic construct, which in our formalization corresponds to the name of a {\sc Coq} file.

In the rest this paper we prefer to use the term \emph{component} instead of \emph{construct}, to emphasize we do not only refer to syntax when we use the term component, but also to its semantics and properties that it may be equipped with. For the semantics of each component to be language-independent, it is necessary that it does not depend on 1) auxiliary entities that are not mentioned by the component, 2) the transition relation of the full language, and 3) abstract syntax of the full language. In our formalization we parametrize the components on these pieces of information. However, we first review MSOS, the framework our formalization is based on.

\begin{table}
\begin{center}
\begin{tabular*}{0.9\textwidth}{l}
\hline
Syntactic Categories\\
\hline
\begin{tabular}{ll}
$\Cmd$ & commands\\
$\Exp$ & expressions\\
$\Dcl$ & declarations\\
{\bf Pcd} & procedure abstractions\\
{\bf Prm} & parameter patterns, encapsulating declarations
\end{tabular}
\end{tabular*}
\par\medskip\noindent
\begin{tabular*}{0.9\textwidth}{l}
\hline
Constructs\\
\hline
\begin{tabular*}{0.9\textwidth}{l @{}l p{0.5\textwidth}}
$\Cmd$ & $::= \on{seq}\;(\Cmd,\,\ldots,\,\Cmd)$ & normal command sequencing\\
$\Cmd$ & $::= \on{skip}$ & normal termination\\
$\Cmd$ & $::= \on{cond-loop}\;(\Exp,\,\Cmd)$ & a simple while-loop, propagating abrupt termination\\
$\Cmd\ $ & $::= \on{catch}\;(\Cmd,\,\Pcd)$ & tries to handle abrupt termination of $\Cmd$ by procedure abstraction $\Pcd$\\
$\Cmd$ & $::= \on{throw}\;\Exp$ & terminates abruptly with the value of the $\Exp$\\
$\Pcd$ & $::= \on{abs}\;(\Prm,\;\Cmd)$ & a parametrized procedure abstraction (with static scoping)\\
$\Prm$ & $::= \on{eq}\;\Exp$ & a parameter that matches only the entity computed by the $\Exp$.\\
$\Exp$ & $::= \on{block}\;(\Dcl,\,\Exp)$ & locally binds $\Dcl$ in the $\Exp$\\
\end{tabular*}
\end{tabular*}
\end{center}
\caption{A basic repository.}
\label{fig:universal}
\end{table}


\subsection{Modular SOS}\label{sec:msos}

In SOS, the operational semantics of a language with effects is modeled in by a \emph{labeled transition system (LTS)} $\langle \Gamma, A, \rightarrow \rangle$, where $\Gamma$ is the set of configurations, $A$ is the \emph{set of actions}, and $\rightarrow\ \subseteq\ \Gamma \times A \times \Gamma$ is the \emph{transition relation} (sometimes called \emph{step relation}). It is possible to consider more general transition systems that include terminal states, but these are only relevant when one considers computation traces, which is outside the scope of this paper. A straightforward example of a set of configurations that we will use below is $\Cmd \times \rho \times \sigma$. We will call $\rho$ and $\sigma$ \emph{auxiliary entities}, or simply \emph{entities}.

A drawback of SOS is its lack of support for modularity. It is sometimes necessary to update existing rules by decorating the transitions with additional entities, e.g. a second store to model a separate part of memory. If we were to add an auxiliary entity to the configurations, then this entity needs to be threaded through \emph{all} the rules that define the semantics. This prevents the rules from being reusable, and therefore plain SOS is not a suitable framework for Component-Based Semantics. One can get around this problem \emph{informally}, by implicitly propagating the entities that are not mentioned, by using a convention such as:

\begin{center}
\begin{tabular}{c c c}
$\inferrule
{\rho \vdash \langle c_1, \sigma \rangle \xrightarrow{}
\langle c_1', \sigma' \rangle}
{\rho \vdash \langle \seq{c_1}{c_2}, \sigma \rangle \xrightarrow{}
\langle \seq{c_1'}{c_2}, \sigma' \rangle}$
&
$\leadsto$
&
$\inferrule{\LT{c_1}{}{c_1'}}{\LT{\seq{c_1}{c_2}}{}{\seq{c_1'}{c_2}}}$
\end{tabular}
\end{center}

\noindent
Normal command sequencing does not manipulate any of the entities and we can therefore assume that they are propagated. This informal description style enables formulation of rules independent of the auxiliary entities that may or may not be present and thereby provides reusability of the rules.

MSOS is a variant of SOS that has special support for the propagation of unmentioned entities. The key distinction is that it separates phrases of the language from entities by moving the entities into a label on the transition. That is, transitions are of the form $\gamma \xrightarrow{\alpha} \gamma'$, such that $\gamma$ and $\gamma'$ merely consist of abstract syntax (which may include computed values), and $\alpha$ is a label containing the auxiliary entities.
Before we discuss the associated transition systems, let us consider some examples of rules specified in MSOS. Figures~\ref{fig:seq} and~\ref{fig:let-in} provide examples of normal command sequencing and local bindings. The abstract syntax is standard, and the meta-variables $c, d, e, \rho$ and $v$ stand for commands, declarations, expressions, environments and values, respectively.

\begin{figure}
 \centering
 \parbox[b]{0.45\linewidth}{
  \centering
  \fbox{Label $:= \{ \ldots \}$}
  \begin{gather}
   \inferrule{}{\LT{\seq{\onskip}{c}}{}{c}\label{rule:seq1}}\\
   \inferrule{\LT{c_1}{\{X\}}{c_1'}}{\LT{\seq{c_1}{c_2}}{\{X\}}{\seq{c_1'}{c_2}}}\label{rule:seq2}
  \end{gather}
  \caption{Normal command sequencing}
  \label{fig:seq}
 }
 \qquad
 \begin{minipage}[b]{0.45\linewidth}
  \centering
  \fbox{Label $:= \{ \rho: \textnormal{env},\,\ldots \}$}
  \begin{gather}
   \inferrule{\LT{d}{\{X\}}{d'}}{\LT{\letin{d}{e}}{\{X\}}{\letin{d'}{e}}}\label{rule:letin1}\\
   \inferrule{\LT{e}{\{\rho=\rho_0[\rho_1],\,X\}}{e'}}{\LT{\letin{\rho_1}{e}}{\{\rho=\rho_0,\,X\}}{\letin{\rho_1}{e'}}}\label{rule:letin2}\\
   \LT{\letin{\rho_1}{v}}{}{v}\label{rule:letin3}
  \end{gather}
  \caption{Local bindings}
  \label{fig:let-in}
 \end{minipage}
\end{figure}

The meta-variable $X$ plays an important r\^ole in the rules. It binds the unmentioned entities, allowing us to propagate them between the premise(s) and conclusion of each rule, without specifically describing what these entities are. Different occurrences of $X$ in the same rule stand for the same entities. Note that the rules assume neither the presence or absence of particular auxiliary entities: the only entities that are mentioned are the ones used by the transitions in the rule in question. The Label box specifies what entities the label should at least include. Entities in labels can be matched in rules using notation such as `$\{\rho = \rho_0[\rho_1],\,X\}$', where $\rho_0[\rho_1]$ stands for updating $\rho_0$ by $\rho_1$. Rules without labels on them are \emph{unobservable}, meaning that they implicitly assume that the entities remain unchanged during the transition (e.g. in rule (\ref{rule:seq1})). As an aside, we remark that \on{skip} too is a component: it has an empty label and an empty set of rules.

Mosses~\cite{Mosses04} recognized that the arrows of a category provide an adequate mathematical structure for labels. That is, two consecutive steps are only allowed to be made when their labels are composable, i.e., $\LT{\LT{\gamma}{p \longrightarrow q}{\gamma'}}{r \longrightarrow s}{\gamma''}$ is only allowed if if $q = r$. Hence, the associated transition systems are a triple $\langle \Gamma, A, \rightarrow \rangle$ similar to LTSes, with the difference that $\Gamma$ strictly consists of abstract syntax, and the additional requirement that $A$ are the arrows of a \emph{label category} $\mathbb{A}$. The label category is a product of elementary categories that correspond to the entities, which we will discuss in Section~\ref{sec:labels}. The values of the auxiliary entities are the objects of $\mathbb{A}$. As an example, a simple step with rule~(\ref{rule:seq1}) looks as follows, if the label contains an environment and a store:

\begin{equation}
\LT{\seq{\onskip}{c}}{
\langle \rho,\,\sigma \rangle
\longrightarrow
\langle \rho,\,\sigma \rangle
}{c}\label{eq:teststep}
\end{equation}

\noindent
Identity arrows are used to express unobservability, used in e.g. rule~(\ref{rule:seq1}).

\section{Formalization}\label{sec:formalization}

In Component-Based MSOS, the source configuration $\gamma$ of a transition $\gamma \xrightarrow{\alpha} \gamma'$ plays a special r\^ole. Namely, it determines to which component the rule permitting that particular transition belongs. The formalization defines for each component a so-called local transition relation, which describes the rules for source configurations that belong to that particular component. Provided with the grammar of the full language, we construct the transition relation of the full language by combining the local transition relations. Components may optionally provide proof of a property that it satisfies, which can likewise be combined to build the proof of that property about the full language (if all components satisfy that property). This will be demonstrated in Section~\ref{sec:example}.

We make use of {\sc Coq}'s support for type classes~\cite{SozeauO08} to automatically ``fill in the details'', i.e. combining the components and filling in the parameters to construct the full language. Type classes, however, are not strictly necessary for the formalization. It is possible in our formalization to construct several full languages from the same repository, but it is not possible to create an extension of an existing full language without completely specifying the extended language's grammar.


\subsection{Types for transition relations}\label{sec:types}

The transition relations of labeled transition systems (see Section~\ref{sec:msos}) can be assigned the following type:

\begin{lstlisting}
Step $\Gamma$ A: $\Gamma$ -> A -> $\Gamma$ -> Prop
\end{lstlisting}

\noindent
In other words, they are predicates which takes arguments $\gamma$, $\alpha$ and $\gamma'$ and return an element of \lstinline{Prop} (the built-in sort of propositional types in {\sc Coq}). Just like the labeled transition systems associated with SOS specifications, there is no apparent distinction between syntax and the auxiliary entities.

Following the principles of MSOS, we update the type of \lstinline{Step} to feature arrows of a category as labels on the transitions. \lstinline{Step} now becomes parametric in the full label category $\mathbb{A}$ of the full language (which has a collection \lstinline{O} of objects), resulting in the following type:

\begin{lstlisting}
Step $\Gamma$ O ($\mathbb{A}$: Category O): $\Gamma$ -> Arrows $\mathbb{A}$ -> $\Gamma$ -> Prop
\end{lstlisting}

\noindent
We have to remark that to avoid confusion, we are not following the exact syntax used in our formalization at this point. Moreover, we omit the definition of \lstinline{Category} in this paper, but we elaborate on \lstinline{Arrows} in Section~\ref{sec:labels}.

Component-Based MSOS requires both a modular way to specify the step relation and a modular way to specify the abstract syntax. The component \on{seq} of Figure~\ref{fig:seq} implicitly specifies its own signature, namely the production rule $\Cmd ::= \seq{\Cmd}{\Cmd}$, and specifies two new rules. It also assumes that a syntactical category $\Cmd$ exists, and to be able to define rule (\ref{rule:seq2}), it assumes that a transition relation on $\Cmd$ exists. We therefore parametrize the component (i.e. its local transition relation and lemmas) with $\Gamma$, representing the syntactic category, the full transition relation \lstinline{S} on $\Gamma$, and the component's construct \lstinline{C} (where \lstinline{P} is a type that stands for its parameters, see the next section). Since the components always define the semantics for precisely one construct of the language, we restrict the input configuration to the phrases built by that construct. We call the transition relation of a component a \emph{local step}, to emphasize the difference with a transition relation defined on a full syntactic category.

\begin{lstlisting}
LocalStep $\Gamma$ O ($\mathbb{A}$: Category O) (S: Step $\Gamma$ O $\mathbb{A}$) P (C: Construct P $\Gamma$):
  restr C -> Arrows $\mathbb{A}$ -> $\Gamma$ -> Prop
\end{lstlisting}

To define the full language, it is sufficient to enumerate the components it is built of. This results in a transition relation of type \lstinline{Step} for each syntactic category, which we call a \emph{global} step relation. This is described later on in this section.


\subsection{Grammar}

As a running example, we define a language that consists of just the components $\on{skip}$ and $\on{seq}$ (see Figure~\ref{fig:seq}). Although it is a fairly simple example, it allows us to explain the formalization without having to get ahead too much on labels, which are treated in Section~\ref{sec:labels}. The grammar of our $\on{skip-seq}$ language is straightforwardly encoded by the following inductive type:

\begin{lstlisting}
Inductive Cmd := skip | seq (c1 c2: Cmd).
\end{lstlisting}

Recall from Section~\ref{sec:cbmsos} that each component is parametrized on its abstract construct. The arguments are passed on as an injection-projection pair which we will call \lstinline{Construct}. Injection corresponds to applying a constructor and projection corresponds to pattern matching. \lstinline{Construct} consists of two properties saying that \lstinline{i} and \lstinline{p} are (partial) inverses of each other. This is needed to prove properties about the component.

\begin{lstlisting}
Class Inject P $\Gamma$ := inject: P → $\Gamma$.
Class Project P $\Gamma$ := project: $\Gamma$ -> option P.
Class Construct P $\Gamma$ {i: Inject P $\Gamma$} {p: Project P $\Gamma$} := {
 H_i: forall x: P, p (i x) ≡ Some x;
 H_p: forall $\gamma$: $\Gamma$, match project $\gamma$ with
       | None => True
       | Some x => i x ≡ $\gamma$ end }.
\end{lstlisting}

\noindent
For constructs that take several arguments, such as $\Cmd ::= \seq{\Cmd}{\Cmd}$, the arguments are tupled. The \lstinline{Class} keyword declares the definitions to be type classes. The convenience of type classes is that class fields (such as \lstinline{inject} or \lstinline{project}) may be used without explicitly mentioning which instance of that class should be used. The curly brackets around \lstinline{i} and \lstinline{p} indicate that these arguments are implicit. In this case, these implicit arguments become class constraints, i.e., order to build an instance of \lstinline{Construct}, instances of \lstinline{Inject} and \lstinline{Project} need to be present. For our example language, the corresponding instances are:

\begin{lstlisting}
Instance: Inject unit Cmd := λ _, skip.
Instance: Inject (Cmd*Cmd) Cmd := λ p, let (c1, c2) := p in seq c$_1$ c$_2$.

Instance: Project unit Cmd :=
  λ $\gamma$, match $\gamma$ with
       | skip => Some tt
       | _ => None end.
Instance: Project (Cmd*Cmd) Cmd :=
  λ $\gamma$, match $\gamma$ with
       | seq c$_1$ c$_2$ => Some (c1, c2)
       | _ => None end.

Instance: Construct unit Cmd.
Instance: Construct (Cmd*Cmd) Cmd.
\end{lstlisting}

\noindent
The type class mechanism can be seen at work here: we do not have to specify the arguments \lstinline{i} and \lstinline{p}, for they can be resolved from the signatures. In fact, the manual declaration of these type class instances is straightforward and can be omitted by an augmentation of {\sc Coq}'s type class resolution algorithm, but we skip the details here. The reader may have noted that when the full language has two constructs with the same signature, the type class instance resolution algorithm may fill in the wrong \lstinline{Construct} instance. This is solved in the formalization by adding an argument (i.e. a \lstinline{string}) to \lstinline{Construct}, enabling us to uniquely identify each instance.

Returning to the \lstinline{LocalStep} type, the \lstinline{Construct} argument is actually a class constraint (i.e. it is an implicit argument) in the formalization. In fact, the category and the \lstinline{Step} relation are also class constraints. Some components require the presence of other components. For instance, the component \on{seq} ``imports'' the (very basic) component \on{skip}. To this end, the \lstinline{Skip} construct becomes an additional constraint of $\on{seq}$. This does not interfere with modularity: all other details about the full language remain opaque.


\subsection{Semantics}

A straightforward way to encode transition relations in a theorem prover is by means of an inductive predicate~\cite{Bertot09}. Making the definition inductive guarantees that the only valid transitions are the ones that can be built by its constructors, which correspond to the rules. The encoding of rules is straightforward using nested implications, where universal quantifications are added for variables that occur in the rules. As an example, we give the transition relation for $\on{seq}$:

\begin{lstlisting}
Inductive ls: restr Seq -> Arrows $\mathbb{A}$ -> Cmd -> Prop :=
| seq_1: forall c$_1$ c$_2$ c$_1'$ ar, step c$_1$ ar c$_1'$ -> ls (Seq <- (c1, c2)) ar (i (c$_1'$, c2))
| seq_2: forall c$_2$ ar, unobs ar -> ls (Seq <- (skip tt, c2)) ar c2.
\end{lstlisting}

\noindent
The premise \lstinline{unobs ar} expresses unobservability of the label, i.e., it has to stay unchanged. We have suppressed the class constraints here for readability. That is, \lstinline{ls} requires suitable instances of \lstinline{Category}, \lstinline{Step}, \lstinline{Construct} and \lstinline{Label} (the latter is presented in Section~\ref{sec:labels}). The type \lstinline{restr C} is used to restrict phrases of the full language to ones built by constructor \lstinline{C}. By means of an inductive type with a single constructor, we can ensure that the only way to build an instance of type \lstinline{restr C} is by providing an object of \lstinline{P}:

\begin{lstlisting}
Inductive restr `(C : Construct P $\Gamma$) := restr_cons ($\gamma$: $\Gamma$).
Notation "C <- $\gamma$" := (restr_cons C $\gamma$) (at level 50, left associativity).
\end{lstlisting}

\noindent
The backtick performs implicit generalization: necessary variables to the argument \lstinline{C} are automatically declared as implicit arguments of \lstinline{restr}. Writing e.g. \lstinline{Seq <- (c1, c2)} is similar to applying the ``real'' constructor (e.g. \lstinline{seq c$_1$ c$_2$}), but not exactly the same. One can obtain \lstinline{c1, c2} by straightforward pattern matching on \lstinline{restr_cons}. In contrast, it is only possible obtain \lstinline{c1, c2} from \lstinline{seq c$_1$ c$_2$} by using the elimination principle of \lstinline{Cmd}, which is not available inside the component.

The inductive predicate \lstinline{ls} is made into a type class instance to enable resolution:

\begin{lstlisting}
Instance LS_Seq: LocalStep O := ls.
\end{lstlisting}


\noindent
The semantics of the full language is essentially defined by a case distinction on the constructors of the datatypes. The full step relation is defined as an inductive predicate \lstinline{s} that combines the existing local step relations of the used components into one global step relation. This is done by means of an inductive predicate that has a single constructor. The constructor assumes a \lstinline{localstep} of any of the local transition relations of the syntactic category in question (passing along \lstinline{s} itself), and returns an object of \lstinline{s} (as above, in \lstinline{ls}). The reader interested in the details is referred to the source code.
This construction satisfies equations such as:

\begin{center}
\begin{tabular}{r@{}c@{}l}
\begin{lstlisting}
localize Skip S_Cmd
\end{lstlisting}$\ $
&
\begin{lstlisting}
=
\end{lstlisting}
&
\begin{lstlisting}
LS_Skip
\end{lstlisting}\\
\begin{lstlisting}
localize Seq S_Cmd
\end{lstlisting}$\ $
&
\begin{lstlisting}
=
\end{lstlisting}
&
\begin{lstlisting}
LS_Seq
\end{lstlisting}
\end{tabular}
\end{center}

\noindent
The operator \lstinline{localize} maps the given \lstinline{Step} instance (in this case \lstinline{S_Cmd}) to the canonical \lstinline{LocalStep} w.r.t. the provided construct. These equations are necessary to prove properties about the components. For example, consider the component $\on{seq}$, which imports the component $\on{skip}$. To be able to prove properties about $\on{seq}$, the local step relation of $\on{skip}$ (which is empty) needs to be accessible. This is done by passing on the first equation as an argument. The equality is overloaded with the obvious meaning that the \lstinline{Step} instances agree on all inputs (i.e. \lstinline{ar}, $\gamma$ and $\gamma'$). In conjunction with {\sc Coq}'s built-in support for setoid rewriting (rewriting modulo an equivalence relation), this enables us to perform short proofs for meta theory (used in Section~\ref{sec:example}).

\section{Labels}\label{sec:labels}

Auxiliary entities such as environments and stores in SOS are encapsulated in a label on the transitions in MSOS. In Section~\ref{sec:cbmsos} we have explained that the labels on the transitions have the structure of arrows of a category: the labels of consecutive transitions should be composable. A subtle difference between MSOS and SOS is that the chosen label category may restrict the transition relation specified by the rules, whereas in SOS it is solely the rules that determine this relation. This can be seen by assuming the label category to be a discrete category, i.e., the category with just identity arrows.

Mosses~\cite{Mosses04} has shown that a suitable category is the product $\mathbb{A} \isdef \prod_{i \in I} \mathbb{A}_i$ of elementary categories representing the auxiliary entities. The usual types of entities used in SOS rules are environments, stores and labels, which correspond to read-only, read-write or write-only permissions, respectively. In MSOS, each entity (with index $i$) has a corresponding set of objects $S_i$ that, together with the permissions, determines its corresponding category $\mathbb{A}_i$:

\begin{itemize}
\item read-only: $\mathbb{A}_i$ is the discrete category with $S_i$ as its objects;
\item read-write: $\mathbb{A}_i$ is the pre-order category with $S_i$ as its objects, and $S_i^2$ as its morphisms;
\item write-only: $\mathbb{A}_i$ is the category with a single object $*$, and the free monoid on $S_i$ as its morphisms.
\end{itemize}

\noindent
A distinguishing feature of MSOS is its inherent support for write-only entities. For example, a transition in a system with a single write-only entity can be pictured as $\gamma \xrightarrow{*\,\longrightarrow\,*} \gamma'$. If it appears as the conclusion of a rule, then the premises of that rule can not possibly depend on the value of that entity, because it is simply $*$. For this reason, we have adopted the use of arrows as labels in our formalization. An alternative is to consider a relation on a product of entities as the label category. This is a special case that does not provide true support for write-only entities.

Recall that the components are parametrized by a label category $\mathbb{A}$ on a collection of objects \lstinline{O}. To build the product category, \lstinline{O} is instantiated with the entity map $i \mapsto \mathbb{A}_i$. Inside the component, the label category is entirely opaque. In other words, it is impossible to learn anything from $\mathbb{A}$ except that it is a product category. The Label box in
the component specification expresses what entities the full label should \emph{at least} include. For example, Figure~\ref{fig:let-in} requires that the full label includes an environment entity. This is reflected in our formalization by providing two functors $\mathsf{P}_M$ and $\mathsf{P}_U$ to each component, that project full labels to their mentioned entities and unmentioned entities, respectively:

\[ \mathbb{A} \stackrel{\mathsf{P}_M}{\longrightarrow} \prod_{i \in M} \mathbb{A}_i,
\qquad \mathbb{A} \stackrel{\mathsf{P}_U}{\longrightarrow} \mathbb{U}. \]

\noindent
The idea is that the product of mentioned entities is transparent to the component, whereas $\mathbb{U}$ is opaque. We use the functor $\mathsf{P}_U$ to express unobservability, needed e.g. in rule (\ref{rule:seq1}). Additionally, the component requires that $(\mathsf{P}_M, \mathsf{P}_U)$ is an isomorphism, which is crucial to enable modular proof. Let us consider determinism as an illustration of this.

\begin{property}
Assume configurations \lstinline{$\gamma$ $\gamma'$ $\gamma''$: $\Gamma$} and labels \lstinline{ar$'$: x ⟶ y}, \lstinline{ar$''$: x ⟶ z}. The step relation on $\Gamma$ is \emph{deterministic} when both
$\gamma\ \xrightarrow{\mathsf{ar'}}\ \gamma'$ and
$\gamma\ \xrightarrow{\mathsf{ar''}}\ \gamma''$ imply that
$\gamma' = \gamma''$ and \lstinline{ar$'$ = ar$''$}.
\end{property} 

\noindent
The requirement that the arrows are equivalent ensures not only that the post configurations are equal, but also the outputs through the write-only components are equal. To prove that the component $\on{seq}$ is deterministic, one proceeds by straightforward case analysis on the structure of the input configuration. In the case that it is $\seq{\onskip}{c}$, we have two arrows \lstinline{ar$'$}, \lstinline{ar$''$} such that \lstinline{$\mathsf{P}_U$ ar$'$ = $\mathsf{P}_U$ ar$''$ = $id$}, and \lstinline{$\mathsf{P}_M$ ar$'$ = $\mathsf{P}_M$ ar$''$ = $()$} (the empty tuple). In other components that do have mentioned entities, these projections of $\mathsf{P}_M$ have to be equivalent. Using the isomorphism we can then conclude that \lstinline{ar$'$ = ar$''$}.


\subsection{Formalization of labels}\label{sec:form_labels}

The category theory we have used in our formalization is provided by the {\sc math-classes} library by van der Weegen and Spitters~\cite{SpittersW11}. Their library makes extensive use of a technique called ``unbundling'', which boils down to separating the components of mathematical structures into separate type classes. An example of this are categories. In Section~\ref{sec:types}, we have treated \lstinline{Category} as a record structure containing \lstinline{Arrows} as a field for presentation purposes. However, in the actual formalization, \lstinline{Arrows} is a separate type class:

\begin{lstlisting}
Class Arrows (O: Type): Type := Arrow: O → O → Type.
Infix "⟶" := Arrow (at level 90, right associativity).
\end{lstlisting}

\noindent
To build a \lstinline{Category}, among other components, an equivalence relation on the corresponding \lstinline{Arrows} instance is necessary, to enable the comparison of arrows. We use this relation in our formalization to define the predicate \lstinline{unobs} for unobservability. The following instances are used for the entity categories:

\begin{lstlisting}
Instance arrows_ro: Arrows O := λ x y, x ≡ y.
Instance arrows_rw: Arrows O := λ x y, unit.
Instance arrows_wo: Arrows unit := λ x y, list O.
\end{lstlisting}

We now define the type class \lstinline{Label}, which is used to provide the projection functors. \lstinline{Label} assumes the presence of the following objects:

\begin{lstlisting}
I M: Type
O: I -> Type
A: forall i: I, Arrows (O i)
O_M: M -> Type
A_M: forall i: M, Arrows (O_M i)
\end{lstlisting}

\noindent
In other words, for both index sets \lstinline{I} and \lstinline{M} it is required that a collection of arrows exists.

\begin{lstlisting}
Class Label := {
 cover_O: forall i: M, O_M i ≡ O (to_I i);
 cover_A: forall i: M, A_M i ≡ $\langle\!\!\langle$ λ T, Arrows T $|$ eq_sym (cover_O i) $\rangle\!\!\rangle$ A (to_I i) }.
\end{lstlisting}

\noindent
The \lstinline{cover_O} property says that for every index of the mentioned entities, the objects have to correspond to the objects of the full category. Likewise, the arrows of the mentioned entities have to correspond. A cast operation~\cite{Hur10} on the objects (indicated by $\langle\!\!\langle\,\_\,|\,\_\,\rangle\!\!\rangle$) is needed to be able to express the latter, but we omit the details in this paper. Given an instance of \lstinline{Label}, we can derive the functors $\mathsf{P}_M$ and $\mathsf{P}_U$ together with the fact that they are isomorphic. Each component has a \lstinline{Label} type class constraint which leaves \lstinline{O} and \lstinline{A} parametric, but specifies \lstinline{O_M} and \lstinline{A_M}.

To illustrate how a rule is interpreted with help of the \lstinline{Label} construction, we consider rule (\ref{rule:letin2}) of Figure \ref{fig:let-in}. Let us first write it using informal notation. Assume that \lstinline{ar: x ⟶ y}, \lstinline{ar$'$: x$'$ ⟶ y$'$} and \lstinline{proj$_\rho$ = $\pi_\rho \circ$ P_M$\!\!$} is the projection of the component with index $\rho$.

\[
\inferrule{
\mathsf{proj}_\rho\ \mathsf{x'} = \rho_0[\rho_1]\\
\mathsf{proj}_\rho\ \mathsf{x} = \rho_0\\
\mathsf{P}_U\ \mathsf{ar} = \mathsf{P}_U\ \mathsf{ar'}\\
\LT{e}{\mathsf{ar'}}{e'}
}
{\LT{\letin{\rho_1}{e}}{\mathsf{ar}}{\letin{\rho_1}{e'}}}
\]

\noindent
In {\sc Coq}-syntax, this rule is:

\begin{lstlisting}
rule4:
  forall ($\rho_0$ $\rho_1$: Env) (e e$'$: Exp) `(ar$'$: x$'$ ⟶ y$'$),
    proj$_\rho$ x$'$ ≡ update $\rho_0$ $\rho_1$  ->  proj$_\rho$ x ≡ $\rho_0$  ->
    fmap_P$_U$ ar =~= fmap_P$_U$ ar$'$  ->  step ar$'$ e e$'$  ->
(* ------------------------------------------------ *)
    ls (Block <- ($\rho_1$, e)) ar (i ($\rho_1$, e$'$))
\end{lstlisting}

\noindent
Note that the use of equality in the above code is highly overloaded, which is made possible by the use of type classes. Like the {\sc math-classes} library, we represent the functors by means of a function that maps the objects, which have the actual names $\mathsf{P}_M$ and $\mathsf{P}_U$, and functions that map the arrows, which have the \lstinline{fmap_} prefix.

\section{Example of Modular Proof}\label{sec:example}

Once the full language is declared, it is possible to combine proofs of the components to prove that a particular property holds for the full language. Like the local step relations, properties are parametrized by a global step relation \lstinline{S}. We say that a property holds for a step relation if it holds for all the possible configurations, but we are a bit more general and allow the user to express that a property holds for a particular configuration.

Not all properties can be proved by induction, and likewise not all properties have a modular proof. We consider a class of admissible, well-behaved properties \lstinline{P} such that \lstinline{P S (I $\gamma$)} does not depend on anything but the localized version of \lstinline{S} w.r.t. \lstinline{C} (here \lstinline{I $\gamma$} injects $\gamma$ into $\Gamma$):

\begin{lstlisting}
Definition admissible $\Gamma$ O (P: Step $\Gamma$ O $\mathbb{A}$ -> $\Gamma$ -> Prop) :=
  forall `(C: Construct A $\Gamma$) (S: Step $\Gamma$ O $\mathbb{A}$) ($\gamma$: restr C),
    P (globalize (localize C S)) (I $\gamma$) -> P S (I $\gamma$).
\end{lstlisting}

\noindent
The operator \lstinline{globalize} is the inverse of \lstinline{localize}: it takes a local step relation \lstinline{ls} and makes it global, behaving like \lstinline{ls} on phrases constructed by \lstinline{C} and not permitting any steps to be made that start from other configurations. The idea of admissible properties is that they warrant that proof by induction is possible.

\begin{lemma}\label{lemma:det_is_adm}
Determinism is admissible.
\end{lemma}

We will demonstrate how this lemma is used to show that our $\on{skip-seq}$ language is deterministic by illustrating the $\on{seq}$ case ($\on{skip}$ is similar). Inside the {\sc Coq} section of $\on{seq}$, we have proved the following lemma that says that the component is deterministic.

\begin{lstlisting}
Lemma det_Seq (c1 c2: Cmd): det_global S_Cmd c$_1$ -> det_local LS_Seq (Seq <- (c1, c2)).
\end{lstlisting}

\noindent
Note that it assumes that the global step relation is deterministic on \lstinline{c1}, which is essentially the induction hypothesis.

Recall the equivalence relations on \lstinline{Step}, \lstinline{LocalStep} of Section~\ref{sec:formalization}. Both \lstinline{det_global} and \lstinline{globalize} respect this relation. Using {\sc Coq}'s built-in support for rewriting modulo equivalence relations (called setoid rewriting), it can be shown that:

\medskip
\begin{tabular}{rll}
&
\begin{lstlisting}
det_global S_Cmd (seq c$_1$ c$_2$)
\end{lstlisting}
& (fold \textsf{I})\\
= &
\begin{lstlisting}
det_global S_Cmd (I (Seq <- (c1, c2)))
\end{lstlisting}
& (Lemma~\ref{lemma:det_is_adm})\\
= &
\begin{lstlisting}
det_global (globalize (localize Seq S_Cmd)) (I (Seq <- (c1, c2)))
\end{lstlisting}
& (rewrite \textsf{eq\_Seq})\\
= &
\begin{lstlisting}
det_global (globalize LS_Seq) (I (Seq <- (c1, c2)))
\end{lstlisting}
& (fold \textsf{det\_local})\\
= &
\begin{lstlisting}
det_local LS_Seq (Seq <- (c1, c2))
\end{lstlisting}
&
\end{tabular}

\medskip
\noindent
Now, the latter holds because this is a property proved in the component $\on{seq}$. The proof for $\on{seq}$ can therefore be completed by applying \lstinline{det_Seq}, using the equation \lstinline{localize Skip S_Cmd = LS_Skip} and the induction hypothesis.

Other components follow the same prescription. In future work, we want to automate the weaving of local proofs by generalizing the above, and exploiting automated proof search with the help of the type class mechanism in {\sc Coq}. Experiments have already demonstrated that this is feasible, but fragile.

\section{Related Work}\label{sec:related}

A specification language for MSOS, called the MSOS Definition Formalism (MSDF), has been developed by Mosses and Chalub (see~\cite{ChalubB07}). It combines BNF notation with textual representation of MSOS transitions. A large number of basic components have already been identified and specified in MSDF. A tool that translates {\sc ML} and (a part of) {\sc Java} into this repository have been developed by Chalub and Braga~\cite{ChalubB07}, which can be executed in the {\sc Maude} tool. MSDF provides its own specification language for datatypes, which can be constructed from primitives such as sequences, lists, maps, etc. In contrast, our formalization directly uses types defined in {\sc Coq}.

Implicit-MSOS is an improvement of MSOS that reduces the amount of clutter in the rules even further by implicitly propagating unmentioned entities~\cite{MossesN09}. The interpretation of Implicit-MSOS is given in terms of MSOS, and we expect that it can be built on top of our formalization by clever use of type classes.

Delaware et al.~\cite{Delaware} have very recently investigated the possibility of modular metatheory in {\sc Coq}. Their focus is on extending a programming language with new features, taking Featherweight {\sc Java} as an example. In their paper, they demonstrate how to develop a modular proof of type-safety of a number of concrete extensions of Featherweight {\sc Java}. The considered extensions do not have effects, i.e., there are no entities.

The formalization of the operational semantics of \textsc{Ocaml}$_\mathrm{light}$ in {\sc HOL} by Scott Owens makes use of labels to encode mutations to the store in them~\cite{Owens08}. These mutations are correlated to a reduction in the program. The labels explicitly carry mutations and therefore simplify the notation, but do not enable a high degree of reusability of the rules.

In a theorem prover (and functional languages), abstract syntax and transition relations are typically encoded as inductive types, of which the constructors correspond to the grammar production rules and the constructors correspond to the rules of the step relation. The inductive definition ensures that those constructors are the only way to build instances of those types. This corresponds to the notions ``initial algebra'' and ``least relation'', sometimes used in this context~(e.g. \cite{MossesN09}). To facilitate Component-Based Semantics, we have to be able to build these inductive types from ``partial versions'' that define just the rules and production rules of the component in question. To our best knowledge, there is no theorem prover (or functional language) that supports (multiple) inheritance of inductive types natively.

\section{Conclusions and Future Work}\label{sec:conclusions}

In this paper we have presented a formalization of Component-Based Semantics in the theorem prover {\sc Coq}. The formalization makes essential use of dependent types, and profits from {\sc Coq}'s support for type classes. Our formalization is based on the ideas of MSOS, and makes use of the idea of labels as arrows in categories, as proposed by Mosses~\cite{Mosses04}. Splitting the label category into a transparent part for the mentioned entities and an opaque part for the unmentioned entities enables modular proof. We have demonstrated this by crafting a proof of determinism of a mini-language from smaller local proofs provided by the components used.

In future work we plan apply this work with the aim of scalable verification of specific programs. Another direction of research is to investigate whether the full generality of labels as arrows (which our formalization provides) can be exploited for entities of types other than read-only, read-write and write-only. We expect that by choosing a suitable category, it is possible to enforce information flow policies, which has applications to security. Our work also enables formal investigation of the appropriate definitions of bisimulation in MSOS, which as of now has an experimental status~\cite{Mosses04}.

\paragraph{Acknowledgments.} The authors wish to thank Peter D. Mosses for introducing them to the notion of Component-Based Semantics, and Bas Spitters for introducing them to type classes in {\sc Coq}. The authors would also like to thank Peter D. Mosses, Julien Schmaltz and the anonymous reviewers for their comments on an earlier version of this paper.

\nocite{*}
\bibliographystyle{eptcs}
\bibliography{fmos}

\end{document}